\documentclass[twocolumn,showpacs,preprintnumbers,amsmath,amssymb]{revtex4}

\usepackage{graphicx,dcolumn,bm,times}

\begin{document}

\title{Impacts of preference and geography on epidemic spreading}

\author{Xin-Jian Xu,$^{1}$ Xun Zhang,$^{2}$ and J. F. F. Mendes$^{1}$\footnote{Electronic address: jfmendes@fis.ua.pt}}

\affiliation{$^{1}$Departamento de F\'{i}sica da Universidade de
Aveiro, 3810-193 Aveiro, Portugal\\
$^{2}$Institute of Theoretical Physics, Lanzhou University, Lanzhou
Gansu 730000, China}

\date{\today}

\begin{abstract}
We investigate the standard susceptible-infected-susceptible model
on a random network to study the effects of preference and geography
on diseases spreading. The network grows by introducing one random
node with $m$ links on a Euclidean space at unit time. The
probability of a new node $i$ linking to a node $j$ with degree
$k_j$ at distance $d_{ij}$ from node $i$ is proportional to
$k_{j}^{A}/d_{ij}^{B}$, where $A$ and $B$ are positive constants
governing preferential attachment and the cost of the node-node
distance. In the case of $A=0$, we recover the usual epidemic
behavior with a critical threshold below which diseases eventually
die out. Whereas for $B=0$, the critical behavior is absent only in
the condition $A=1$. While both ingredients are proposed
simultaneously, the network becomes robust to infection for larger
$A$ and smaller $B$.
\end{abstract}


\maketitle

The classical mathematical approach to diseases spreading either
ignores the population structure or treats populations as
distributed in a regular medium \cite{Bailey_75,Anderson_85}.
However, it has been suggested recently that many social,
biological, and communication systems possess two universal
characters, the small-world effect \cite{Watts_98} and the
scale-free property \cite{Barabasi_99}, which can be described by
complex networks whose nodes represent individuals and links
represent interactions among them \cite{Dorogovtsev_03,Pastor_04}.
In view of the wide occurrence of complex networks in nature, it is
important to study the effects of topological structures on the
dynamics of epidemic spreading. Pioneering works
\cite{Diekmann_90,Anderson_92,Moore_00,Lloyd_01,Pastor_01} have
given some valuable insights: for homogeneous networks (e.g.,
exponential networks), there are critical thresholds of the
spreading rate below which infectious diseases will eventually die
out; on the contrary, even infections with low spreading rates will
prevail over the entire population in heterogeneous networks (e.g.,
scale-free networks). This radically changes many conclusions drawn
from classic epidemic modelling. Furthermore, it has been observed
that the heterogeneity of a population network in which the disease
spreads may have noticeable effects on the evolution of the epidemic
as well as the corresponding immunization strategies
\cite{Albert_00,Moreno_02,Pastor_02,Eames_03,Cohen_03}.

For many real networks, however, individuals are embedded in a
Euclidean space and the interactions among them usually depend on
their spatial distances and take place among their nearest neighbors
\cite{Durrett_99,Yook_02,Gastner_06}. For instance, the number of
long-ranged links and the number of edges connected to a single node
are limited by the spatial embedding, particularly in planar
networks. Preferential attachment is weakened by geographical
embedding \cite{Yook_02}. Also, people have proved that the
characteristic distance plays a crucial role in the dynamics taking
place on those networks
\cite{Rozenfeld_02,Warren_02,Manna_04,Mukherjee_06}. Thus, it is
natural to study associated influences of preference and geography
on epidemic spreading. But up to now only a few of works address
this problem, e.g., modeling transmission as a function of
geographical distance \cite{Durrett_94,Hethcote_00} availably
capturing the dynamics of diseases in wild and domesticated animals
\cite{Murray_86,Keeling_03}.

In this paper, we study the standard
susceptible-infected-susceptible (SIS) model on a growing network in
Euclidean space. On a vertical plane, the growth of the network
depends jointly on two mechanisms, preference and geography. The
placement of links is driven by competition between preferential
attachment and distance dependence. In the case that the network
grows with geographical constraint, we recover the usual epidemic
behavior with a critical threshold below which diseases will
eventually die out. While the network is totally governed by
preferential attachment, the epidemic behavior depends on the
preferential exponent. When both factors are considered
simultaneously, it becomes difficult for epidemic spreads as the
preference has an overwhelming majority than the geography.

\begin{figure}
\includegraphics[width=\columnwidth]{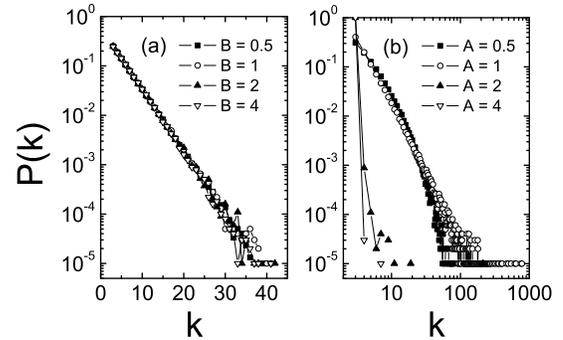}

\caption{Degree distributions of the generated networks with
$m_{0}=m=3$ in two special cases,  $A=0$ (a) or $B=0$ (b). The size
of the network is $N=10^{5}$.} \label{fig1}
\end{figure}

Specifically in two-dimensional $x-y$ plane, we consider a square of
unit size and with periodic boundary conditions. To construct a
network of $N$ nodes, let $(x_1,x_2,\dots,x_N)$ and
$(y_1,y_2,\dots,y_N)$ be the $2N$ independent random variables
identically and uniformly distributed within the interval $\{0,1\}$.
A specific set of values of the random variables
$\{(x_1,y_1),(x_2,y_2),\dots,(x_N,y_N)\}$ is chosen to represent the
coordinates of the $N$ randomly distributed nodes \cite
{Mukherjee_06}. The network starts with $m_{0}$ nodes and then the
other nodes with $m$ links are added one by one at each time step
according to their serial numbers $i=m_0+1$ to $N$. Following ideas
proposed by Yook et al. \cite {Yook_02}, the probability that a new
node $i$ links to a old node $j$ with $k_j$ links at distance
$d_{ij}$ from node $i$ is
\begin {equation}
\prod (k_{i},d_{ij}) \sim \frac{k^{A}_{j}}{d^{B}_{ij}},
\label{linkprod}
\end {equation}
where $A$ and $B$ are positive constants, governing preferential
attachment and the cost of the node-node distance. We note following
interesting features. (i) In the case of $A=0$, the network is
geographically grown with an exponential distribution of nodes'
degree (see Fig. \ref{fig1}(a)). In the limit of $B \to \infty$,
only the smallest value of $d$ corresponding to the nearest node
will contribute with probability $1$ \cite {Yook_02}. (ii) In the
case of $B=0$, the network reduces to the Barab\'{a}si-Albert (BA)
graph only for $A=1$. In the region $0<A<1$, the nodes' degree
distribution is stretched exponential. For $A>1$, a finite number of
nodes connect to nearly all other nodes \cite {Krapivsky_00}. That
is illustrated by Fig. \ref{fig1}(b).

To estimate the effect of the network's topology on epidemic
dynamics, we will investigate the standard SIS model \cite
{Anderson_92}. This model relies on a coarse-grained description of
individuals in the population. Namely, each node of the network
represents an individual and each link is a connection along that
the infection can spread to other individuals. The individuals can
only exist in two states, susceptible and infected. At each time
step, each susceptible node is infected with probability $\nu$ if it
is connected to one or more infected nodes. At the same time, the
infected nodes become susceptible again with probability $\delta$,
defining an effective spreading rate $\lambda=\nu/\delta$. We can
set $\delta=1$ without lack of generality, since it only affects the
definition of the time scale of the virus propagation. Individuals
run stochastically through the cycle susceptible $\rightarrow$
infected $\rightarrow$ susceptible.

\begin{table}
\caption{The average path length $L$ and clustering coefficient $C$
of the generated networks with size $N=10000$ in the case of $A=0$
for different values of $B$.} \label{tab1}
\begin{ruledtabular}
\begin{tabular}{ccc}
$B$ & $L$ & $C$\\
\hline
0.5 & 5.13(1) & 0.00(1)\\
1 & 5.13(5) & 0.00(2)\\
2 & 5.40(1) & 0.03(9)\\
3 & 6.41(7) & 0.19(2)\\
4 & 7.11(2) & 0.31(0)\\
\end{tabular}
\end{ruledtabular}
\end{table}

\begin{figure}
\includegraphics[width=\columnwidth]{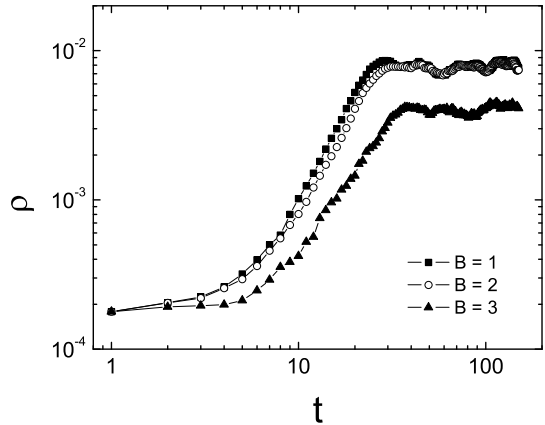}

\caption{Density of infected nodes $\rho$ as a function of time $t$
in the network with $A=0$ and $B=1,2, 3$, respectively. The
spreading rate is $\lambda=0.15$. Simulations were performed on
networks with $N=10000$ and $m=m_{0}=3$.} \label{fig2}
\end{figure}

Let us focus on the case $A=0$ first, i.e., the network grows with
geographical constraint. According to Eq. (\ref{linkprod}), the
preferential attachment is excluded. Since all nodes are uniformly
distributed in the square, the only effect of the factor $B$ is
determining the average path length of the network while the degree
distribution of nodes has the same behavior. For small $B$, the role
of node-node distance is weak and old nodes are linked with
approximate randomness. When $B$ becomes large, the geographical
influence is strong and only nodes around the new one will be
connected with large possibility, hence the local clustering. This
feature is reflected in Tab. \ref{tab1}, that is, the average path
length $L$ and the clustering coefficient $C$ get larger with the
increase of $B$.

\begin{figure}
\includegraphics[width=\columnwidth]{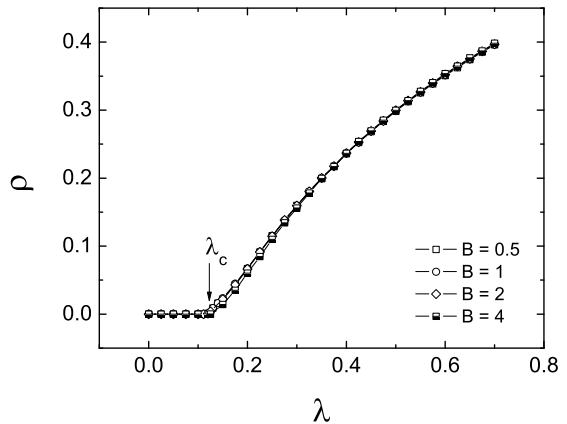}

\caption{Density of infected nodes $\rho$ as a function of $\lambda$
in the network with $A=0$ and $B=0.5,1,2,4$, respectively.}
\label{fig3}
\end{figure}

Figure \ref{fig2} shows the evolution of the infected nodes density
as a function of time for epidemics with $\lambda=0.15$. We start
from a single infected node of the network, and iterate the rules of
the SIS model with parallel updating. Each curve represents the
average over $10$ different starting configurations, performed on
$10$ different realizations of the random networks. We clearly
notice a great influence of the geography on the spreading velocity
of diseases, namely, the smaller the parameter $B$ is, the more fast
the infection propagates. For finite $B$, since the geography does
not change the network's connectivity distribution (see Fig.
\ref{fig1}(a)), the evolution of node i's degree can be written as
\begin{equation}
\frac{d k_i(t)}{d t} \approx \frac{m}{t},
\end{equation}
with the initial condition $k_i(i)=m$. One can easily write the
solution
\begin{equation}
k_i(t)=m(\ln\frac{t}{i}+1),
\end{equation}
and accordingly obtain the degree distribution
\begin{equation}
P(k)=\frac{1}{t}\sum_{i=1}^{t}\delta(k_i(t)-k)=\frac{1}{m}e^{-\frac{k}{m}+1}.
\label{degreedistribution}
\end{equation}
In complex networks, the basic reproductive number takes the form,
$R_0 \sim \langle k^2 \rangle / \langle k \rangle$ \cite
{Moreno_02}. Different from the classical result, it defines an
epidemic threshold $\lambda_c = \langle k \rangle / \langle k^2
\rangle$. Combining Eq. (\ref{degreedistribution}), we have
\begin{equation}
\lambda_c = \frac{\int k\frac{1}{m}e^{-\frac{k}{m}+1}dk}{\int
k^{2}\frac{1}{m}e^{-\frac{k}{m}+1}dk} = \frac{2}{5m}.
\label{lambdac}
\end{equation}
In Fig. \ref{fig3}, we plot the steady density of infected nodes
$\rho$ as a function of the spreading rate $\lambda$ for the case of
$A=0$, which is the time average of the fraction of infected
individuals reached after an initial transient regime. Simulations
were computed over $50$ different starting configurations, performed
on $50$ different realizations of the networks. The size of networks
is $N=10^{5}$. As shown in Fig. \ref{fig3}, all curves display the
same behavior and the SIS model exhibits an epidemic threshold,
$\lambda_c = 0.12(2)$, despite the variety of $B$, which is in
agreement with the analytical prediction, $\lambda_c = 2 /(5 \times
3) = 0.133$ (Eq. (\ref{lambdac})) \cite {Note_1}. That is different
from the results gained by Santos et al., who also studied the SIS
model on a homogeneous small-world network \cite{Santos_05}, the
critical value of $\lambda$ changes smoothly as one varies rewiring
probability without changing the degree distribution. In our model,
the increase of $B$ gives rise to the average path length and the
clustering coefficient. Whereas in \cite{Santos_05}, those network
features reduce with the increase of the rewiring probability.

\begin{figure}
\includegraphics[width=\columnwidth]{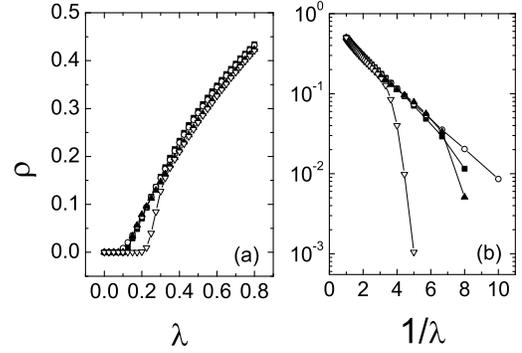}

\caption{Density of infected nodes $\rho$ vs. $\lambda$ (a) and
$1/\lambda$ (b) in the network with $B=0$. The preference exponents
are $A=0.5$ (closed squares), $1$ (open circles), $1.5$ (closed
diamonds), and $2$ (open triangles), respectively.}\label{fig4}
\end{figure}

\begin{figure}
\includegraphics[width=\columnwidth]{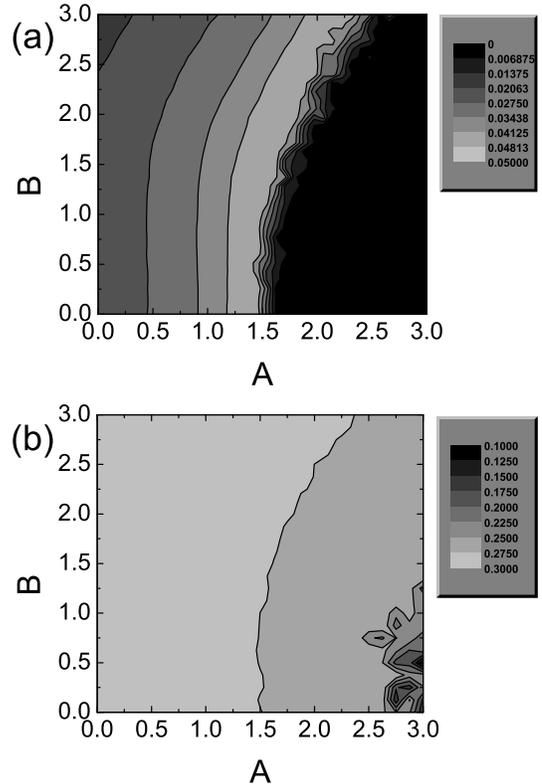}

\caption{Density of infected nodes $\rho$ in $A-B$ plane for the
spreading rate $\lambda=0.15$ (a) and $0.5$ (b),
respectively.}\label{fig5}
\end{figure}

Next, we will investigate the dynamics in the case of $B=0$, i.e.,
the network grows following preferential attachment. For the linear
preference, $A=1$, the generated network reduces to the exact BA
graph. As shown in Fig. \ref{fig4}, the behavior of the density of
infected nodes follows the property, $\rho \sim e^{-1/m\lambda}$
\cite{Pastor_01}, which implies the surprising absence of any
epidemic threshold in the model, i.e., $\lambda_c = 0$. As to the
nonlinear case, we first consider the region $A<1$ and go down a
little from $A=1$. In this event, highly connected nodes become less
attractive for attachment compared with the linear preference. The
resulting degree distribution is of the form \cite{Krapivsky_00},
$P(k) \sim k^{-A}\exp(-\mu k^{1-A}/(1-A))$, where $\mu$ is a
positive constant depends on $A$,
$\mu=\sum_{k=2}^{\infty}\prod_{i=2}^{k}(1+\mu/i^{A})^{-1}$. In the
region $A>1$, the attractiveness of the old highly connected nodes
increases which results in a small number of nodes that get all
connections in the network. As shown in Fig. \ref{fig4}, there are
epidemic thresholds of the SIS model for $A=0.5, 1.5$, and $2$,
respectively. Furthermore, the threshold increases as $A$ becomes
larger.

Finally, we plot the prevalence $\rho$ in $A-B$ plane in Fig.
\ref{fig5} to shown the influence of the competition of two
ingredients on infections. For $\lambda=0.15$, there exists a set of
peaks in Fig. \ref{fig5}(a). Namely, given the value of $B$, as $A$
is increased initially, the density of infected nodes $\rho$
increases gradually and reaches a maximum for some value of $A$, and
then decreases rapidly to $0$ as $A$ is increased further.
Furthermore, the contour planes of $\rho$ takes an excursion to
right with the increase of $B$. We argue that the following factor
should be taken into account to understand this performance.
According to Eq. (\ref{linkprod}), as $B$ increases, the new nodes
are preferential to connect their nearest neighbors, which results
in the decrease of the number of long-ranged links. The network
becomes more local clustering and robust to epidemic spreading. To
keep the same prevalence, the effect of node's degree should be
strengthened, i.e., increases $A$. Thus the contour planes lean to
right as $B$ becomes larger. For $\lambda=0.5$, the prevalence
$\rho$ displays a different behavior. As shown in Fig.
\ref{fig5}(b), $\rho$ decreases monotonically as $A$ gets larger,
and if $B$ decreases at the same time, the network becomes robust to
diseases.

To summarize, we have studied the SIS model on a random network. On
an Euclidean plane, the network grows depending jointly on the
preference and the geography. The former indicates the
attractiveness of highly connected nodes and the later denotes the
geographical constraint. It is found that both factors have great
influences on the infection. For the network growing with the
geographically constraint ($A=0$), we recover the usual epidemic
behavior with a critical threshold below which diseases will
eventually die out. While the network is purely governed by
preferential attachment ($B=0$), the epidemic behavior depends on
the preferential exponent and the critical phenomenon is absent only
in the condition $A=1$. When both factors are present
simultaneously, the network becomes robust to diseases as the
preference has an overwhelming majority than the geography. In real
world, agents located on different positions according to the
competition between the preference and the geography. The above
description of the spreading dynamics might contribute to
understanding realistic epidemics.

This work was partially supported by DYSONET 012911 and FCT
SFRH/BPD/30425/2006.


\begin{references}
\bibitem{Note_1}
In fact, $\lambda_c$ for different $B$ only changes in permillage
and we deem the independence of the critical value on geographical
constraint.
\end{references}
\end{document}